\documentclass[
reprint,
amsmath,amssymb,
aps,
prl,
]{revtex4-2}

\usepackage{float}
\usepackage{mathtools}
\usepackage{graphicx}
\usepackage{dcolumn}
\usepackage{bm}
\usepackage{upgreek}
\usepackage{lipsum, babel, xcolor}
\usepackage{amsmath, amssymb}
\usepackage{comment}
\usepackage{natbib}
\usepackage{physics}
\usepackage{hyperref}
\hypersetup{colorlinks=true,citecolor=blue}

\begin{document}


\title{High bandwidth traveling wave electro-optic modulator at $\mathrm{1\mu m}$ on thin-film lithium tantalate}
\author{Ayed Al Sayem, Shiekh Zia Uddin, Ting-Chen Hu, Alaric Tate, Mark Cappuzzo, Rose Kopf, Mark Earnshaw}

\affiliation{%
  Nokia Bell Labs, NJ, USA 
}%
\date{\today}

\begin{abstract}
We present the first experimental demonstration of a high-bandwidth thin-film lithium tantalate (TFLT) electro-optic modulator operating at 1 $\mu$m, with a $\mathrm{V_{\pi}}$ of $2.4$\,V, and less than $2$\,dB electro-optic roll-off upto $50$\,GHz and stable DC bias operation.
\end{abstract}

\maketitle

\section{Introduction}

The recent advancement of hollow-core fibers now opens up the opportunity to utilize a wide range of wavelengths, such as the visible wavelength band and also the non-conventional near IR wavelengths bands, such as 780\,nm, 850\,nm and 1\,$\mathrm{\mu m}$ with lower optical losses than achievable with standard glass core fiber \cite{Petrovich2025NatPhotonics,Shi2025NatCommun,Gao2025Optica,Wan2024OE,Ge2025OFC,Ge2024OFC,Mukasa2023OFT}. Hollow-core fibers can also operate at much higher optical power due to very weak nonlinearity \cite{Cooper2023HCFHighPower,Shi2025NatCommun,Michieletto2016HCFPulseDelivery}, potentially enabling the operation of the optical transceiver with very high optical power and potentially eliminating or significantly reducing the number of line amplifiers used many communication networks. To take full advantage of hollow-core fibers, one requires a material system with low optical loss, strong electro-optic effect, high power-handling capability, and stable operation. Unfortunately, the most explored and commercially utilized materials, such as Silicon (Si) or Indium phosphide (InP), have a band-gap of $\mathrm{1.1\,\mu m}$ and $\mathrm{0.9\,\mu m}$ respectively, which makes these platforms not ideal to take advantage of the transparency window of the hallow core fibers below the O-band. Among the transparent materials with a wide optical window, lithium niobate (LN) and lithium tantalate (LT) are particularly promising because they offer broadband transparency along with a strong electro-optic effect \cite{zhu2021integrated,wang2024lithium}. The cut-off wavelengths for LN and LT are 350\,nm \cite{zhu2021integrated}, and 316\,nm respectively \cite{wang2024lithium}. Thin-film lithium niobate (TFLN) has already been widely explored in the visible and near-infrared range for applications such as generating efficient second-harmonics \cite{wang2018ultrahigh,Lu2020,sayem2021efficient}, entangled photon pairs \cite{ma2020ultrabright}, optical parametric generation \cite{nehra2022few}, and low-loss resonators \cite{zhang2017monolithic}, and so on. TFLN electro-optic modulators in the visible and near-infrared regimes have been widely explored. For example, TFLN modulators have already been demonstrated at 532\,nm \cite{Li2022OE}, 780\,nm \cite{Celik2022OE}, 850\,nm \cite{Assumpcao2025OL}, and at 1064\,nm \cite{Jagatpal2021LPT}. The fundamental advantage of operating in such a wavelength range comes directly from the tighter mode confinement, which allows closer electrode spacing, which in turn reduces the drive voltage \cite{Assumpcao2025OL,powell2025sub}. The second fundamental advantage originates purely from physics, as the phase accumulation per length is inversely proportional to the operating wavelength. Unfortunately, TFLN suffers significantly from the well-known photo-refractive (PR) effect \cite{xu2021mitigating,Ahmed2025Universal,ren2025photorefractive,Li2019PhotonLevelTuning,arge2025demonstration}. Because of the PR effect, TFLN is not DC-bias stable \cite{xu2020high,Powell2024OE_TFLT_DCStable}, requiring high power thermal control, which limits its practical utility for large-scale photonic circuits. Critically, the PR effects are more drastic at visible and near-infrared wavelengths as carriers from defect-related trap centers such as $\mathrm{Fe^{2+}}$, and $\mathrm{Fe^{3+}}$ ions as well as other intrinsic defects can be easily photo-excited at these wavelength bands, which have been extensively studied for bulk LN \cite{Jermann1993LightInducedChargeTransportFeLN,Goulkov2014PhotoelectricResponseFeRatio,Furukawa2001GRIIRA_MgO_LN,Volk2009LithiumNiobateDefectsPhotorefraction}. For a practical communication system, one needs a material system that not only offers high performance but also maintains stable performance. We recently showed that thin-film lithium tantalate (TFLT) is a highly stable material platform for optical power handling and outperforms most material platforms, such as Si, LN, $\mathrm{Ta_{2}O_{5}}$, etc., in the C-band. For short-wavelength operation, we also expect TFLT to be a better choice of material platform than TFLN \cite{powell2025sub,guo2026robust} due to the stronger PR effect at the shorter wavelength bands \cite{desiatov2019ultra}. Unfortunately, TFLT is not as explored as TFLN, and there have been very few experimental demonstrations \cite{wang2024lithium,zhang2025ultrabroadband,wang2024ultrabroadband}, especially at wavelengths beyond the telecommunication band \cite{powell2025sub,guo2026robust}. Here, we show the first experimental demonstration of a high-speed ($\mathrm{>50\,GHz}$ electro-optic bandwidth with $\mathrm{<2\,dB}$ roll-off) modulator with a $\mathrm{V_{\pi}}$ of 2.4\,V on a standard 600\,nm TFLT 4-inch wafer platform with a wafer-level fabrication process flow. We also show that our device is DC-bias stable even at $\mathrm{1\,\mu m}$ and can generate sharp pulses without distortion even with very wide pulses. 

\begin{figure*}[ht]
    \centering
    \includegraphics[width = 0.85\textwidth]{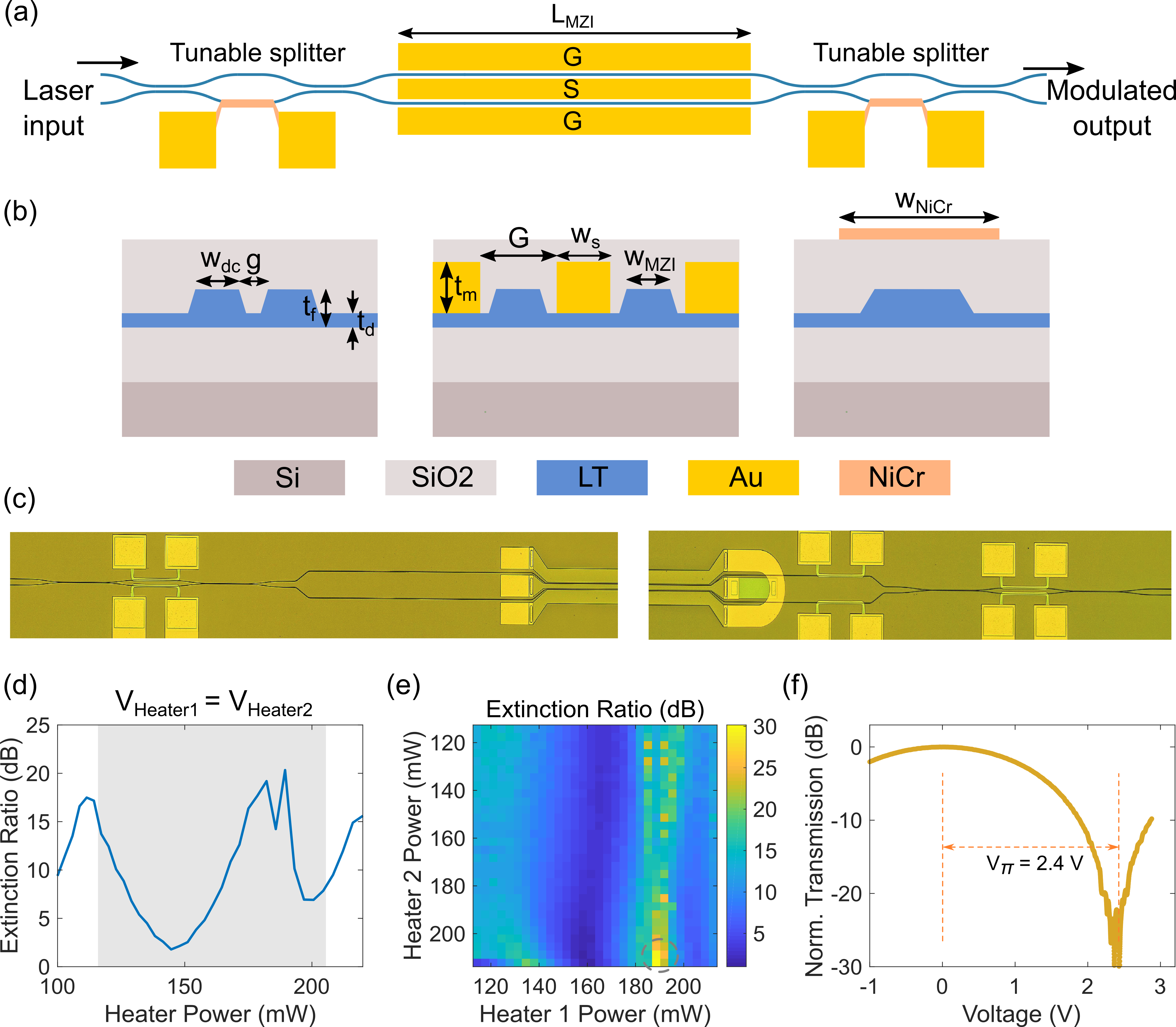}
    \caption{(a) Schematic of the modulator device. G: ground, S: signal (b) Cross-section of the modulator in three different sections of the device, i) the directional coupler (DC), ii) MZI section, iii) waveguide with NiCr heater for thermal tuning. Here, $\mathrm{L_{MZI}=7\,mm,,w_{DC}=0.8\,\mu m, g=0.8\,\mu m, t_{f}=600\,nm, t_{d}=\sim240\,nm, t_{m}=1\,\mu m, w_{s}=17\,\mu m, w_{MZI}=1.6\,\mu m, w_{NiCr}=10\,\mu m.}$ (c) Optical microscope image of the device. (d) Extinction ratio of the modulator as a function of the heater power with identical heating power for two heaters. (e) Two-dimensional (2D) color of extinction as a function of heater power. (f) Normalized transmission of the modulator as a function of applied triangular voltage at $100$ Hz.}
    \label{Fig1}
\end{figure*}
\section{Device geometry}
In Fig.\ref{Fig1}, we show the schematic of the traveling wave electro-optic modulator used in this article. The modulator consists of three cascaded Mach-Zehnder interferometers (MZIs). The first and the third MZI act as tunable splitters so that we can operate the modulator at a wide range of wavelengths from $\mathrm{1.5\,\mu m}$ to near $\mathrm{1\,\mu m}$. Fig.\ref{Fig1}(b) shows the cross-section of different sections of the device, the directional coupler (DC), modulator section, and heater section, respectively. Fig.\ref{Fig1}(c) shows the optical microscope image of the modulator. The device has been fabricated on a 4-inch 600\,nm thick TFLT wafer with $\mathrm{4.7\mu m}$ thick insulating oxide underneath, commercially available from NanoLN. We use on-chip $\mathrm{50\,\Omega}$ termination for traveling-wave operation. Details of the device fabrication can be found in our previous articles \cite{al2025multi,sayem2026high}. By controlling the phase of the first and the third MZI, e.g., the tunable splitters, we can control the beam splitting ratio over a wide range of wavelengths. Fig.\ref{Fig1}(d) shows the extinction ratio of the modulator as a function of the heater power when identical heating power is applied to both heaters. Fig.\ref{Fig1}(e) shows the two-dimensional color map of the extinction ratio when the heater power was swept for both heaters. It can be observed from Fig.\ref{Fig1}(e) that close to 30\,dB extinction can be achieved, which is limited by the dynamic range of the photo-detector (PD) used in the experiment. In Fig.\ref{Fig1}(f), we show the normalized transmission of the device as a function of applied voltage. All these measurements are performed at $\lambda=1071\,nm$.
\begin{figure*}[ht]
    \centering
    \includegraphics[width = 0.80\textwidth]{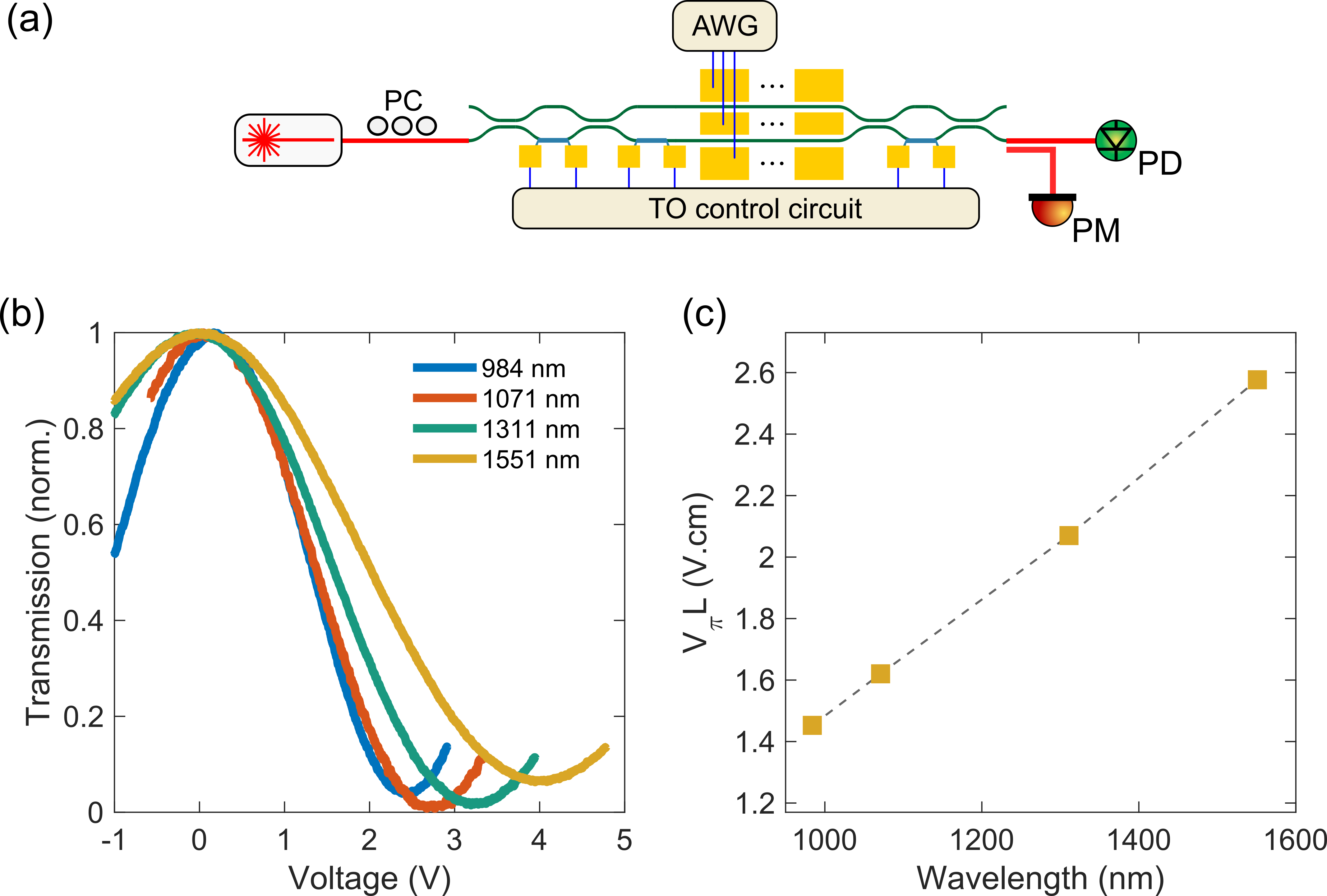}
    \caption{Transmission as a function of applied voltage for four different wavelengths of operation, $\mathrm{\lambda=984\,nm}$, $\mathrm{\lambda=1071\,nm}$, $\mathrm{\lambda=1311\,nm}$, and $\mathrm{\lambda=1551\,nm}$. (b) $\mathrm{V_{\pi}L}$ as a function of operating wavelength.}
    \label{Fig2}
\end{figure*}
\section{Modulator performance}
\begin{figure*}[ht]
    \centering
    \includegraphics[width = 0.85\textwidth]{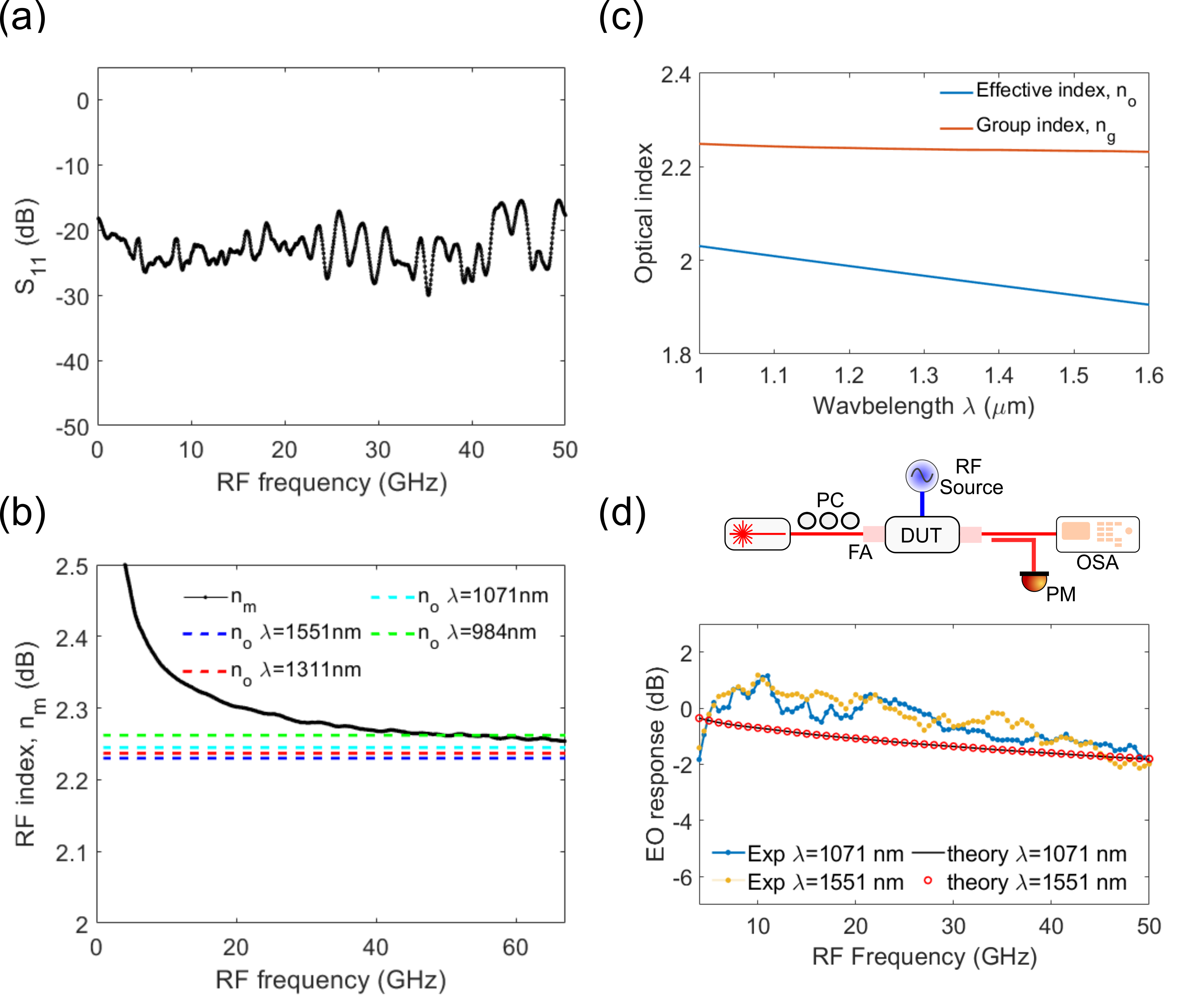}
    \caption{(a) Experimental $\mathrm{S_{11}}$ of the  TFLT modulator. (b) Extracted RF phase index as a function of RF frequency. (c) Simulated optical phase and group index of TFLT waveguides as a function of wavelength. Here, the waveguide width is $\mathrm{1.6\,\mu m}$. (d) Schematic of the measurement setup and electro-optic (EO) response of the modulator as a function of RF frequency at two operating wavelengths.}
    \label{Fig3}
\end{figure*}

We characterize the modulator performance at relevant wavelengths for a systematic comparison. The schematic of the measurement setup is shown in Fig.\ref{Fig2}(a). Light from either a diode laser or a tunable laser is sent to the device under test (DUT) through a fiber-based polarization controller and a lensed fiber. We use different lasers to test the modulator at different wavelengths. For $\mathrm{\lambda=984\,nm}$, we use a fixed wavelength semicondutor laser. For $\mathrm{\lambda=1071\,nm}$ and $\mathrm{\lambda=1551\,nm}$, we use tunable lasers from Santec (Santec-570), and for $\mathrm{\lambda=1311\,nm}$, we use a tunable laser from EXFO (model). The output light from the device is collected by another lensed fiber and is then sent to a slow power-meter (PM) and a fast photo-detector (PD). Fig.\ref{Fig2}(b) shows the normalized transmission as a function of applied voltage for the same modulator at different wavelengths. In Fig.\ref{Fig2}(b), we plot the voltage-length ($\mathrm{V_{\pi}L}$) product as a function of wavelength. Clear wavelength scaling can be observed in Fig.\ref{Fig2}(c). Here, we use the same device with an electrode gap, $\mathrm{G=5\,\mu m}$. For shorter wavelengths, it's possible to use a smaller electrode gap and improve $\mathrm{V_{\pi}}$ ($\mathrm{V_{\pi}L}$) \cite{powell2025sub}. 
We also completed a radio-frequency (RF) characterization of the modulator. In Fig.\ref{Fig3}(a), we plot the RF reflection, $\mathrm{S_{11}}$ of the modulator. We measure the RF transmission spectra of the modulator on a separate device without termination resistors but with identical geometry. In Fig.\ref{Fig3}(b), we show the calculated microwave phase index from the RF transmission measurements along with the optical group index at different wavelengths. In Fig.\ref{Fig3}(c), we plot the simulated optical phase index, $\mathrm{n_{o}}$, and group index, $\mathrm{n_{g}}$ as a function of wavelength for TFLT waveguides for the geometry used in this paper. From Fig.\ref{Fig3}(c), we can observe that TFLT waveguides offer a flatter dispersion, which indicates the possibility of high-bandwidth operation over a wide range of wavelengths simultaneously. Due to the lack of high-speed photo-detectors, we measure the electro-optic (EO) bandwidth of the modulator at $\mathrm{1\,\mu m}$ using an optical spectrum analyzer (OSA). Fig.\ref{Fig3}(d) shows the schematic of the measurement setup and the measured EO response as a function of the RF drive frequency at two different wavelengths, e.g., at 1071\,nm and 1551\,nm. Almost identical EO roll-off can be observed for both wavelengths, proving the ultra-flat dispersion of LT. Less than 2\,dB roll-off can be observed for RF frequency up to 50\,GHz. 
\begin{figure*}[ht]
    \centering
    \includegraphics[width = 0.85\textwidth]{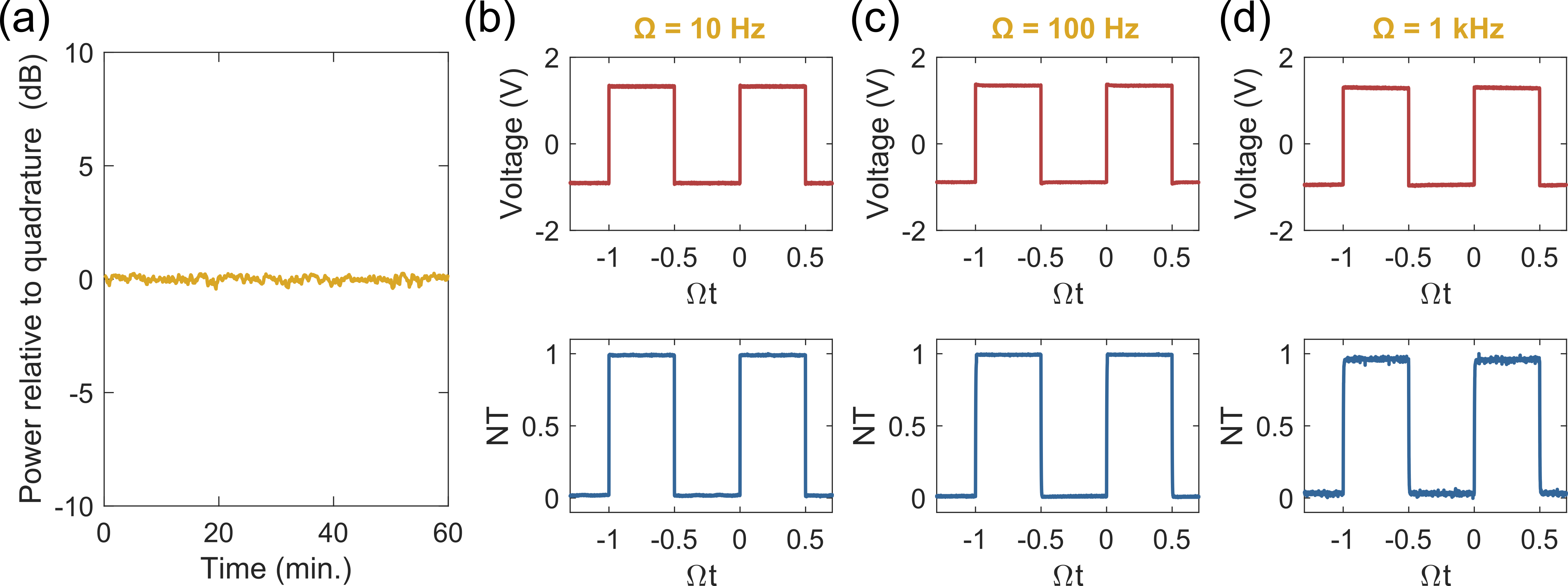}
    \caption{(a) Normalized Transmission (NT) of the MZI modulator as a function of time when the modulator is biased at the quadrature point. Applied voltage and corresponding output pulse from the modulator for square-wave signals at (b) $\Omega = 10$\,Hz, (c) $100$\,Hz, and (d) $1$\,KHz.}
    \label{Fig4}
\end{figure*}

We next investigate the stability of the TFLT modulators near $\mathrm{\lambda=1\,\mu m}$. TFLN is widely known for its DC bias instability \cite{Zhao2025SPIE_DCdift_TFLN,Celik2024BiasDrift,Holzgrafe2024OE_EORelaxation,xu2020high}, and TFLT modulators have already been shown to have much better DC bias stability for both waveguide and resonant modulators \cite{Powell2024OE_TFLT_DCStable,sayem2026high}. Most of these experiments have been performed near the C-band. The PR effect is much stronger at shorter wavelengths, such as in the visible and near-infrared, hence we study the DC-bias stability and pulse generation at $\mathrm{\lambda=1\,\mu m}$. Fig.\ref{Fig4}(a) shows the output power of the TFLT modulator as a function of time when the modulator is biased at the quadrature point. Negligible drift is observed over more than one hour with an on-chip power of  $\mathrm{\sim0\,dBm}$, showing excellent DC bias stability of the modulator. Sharp pulse generation is another key characteristic of a stable modulator. Because of the PR effect and defect-assisted charge dynamics, with TFLN modulators, it's difficult to generate sharp pulses \cite{shen2023ultra}. Fig.\ref{Fig4}(b)–(d) show the applied voltage and corresponding output pulse from the modulator when square-wave signals at 10\,Hz, 100\,Hz, and 1\,kHz, respectively, are sent the modulator. The output waveform faithfully tracks the input waveform, preserving sharp pulse edges without observable distortion, which suggests that charge-related effects are negligible in the device. These results confirm that our modulator can generate sharp pulses at 1\,$\mu$m at low frequencies where charge related distortions are most pronounced, detailed results on achieving high extinction ratio pulses in TFLT devices will be presented in a future work.

\section{Acknowledgment}
We thank Santec Corporation for providing us the TSL-570 at $\mathrm{1\,\mu m}$.

\section{Author contribution}
A.S. designed the photonic devices and developed the fabrication process flow with T.H, A.T, M.C., and R. K.. T.H, A.T, M.C., and R. K. fabricated the devices. A.S and S.U performed the measurements. A.S, S.U wrote the paper with technical feedback from M.E.

\section{Funding}
Nokia Corporation of America.

\bibliography{Reference}


\end{document}